\documentclass[aps,prl,twocolumn,showpacs,a4paper]{revtex4}

\usepackage{graphicx,color}
\usepackage{ifthen}
\usepackage[pdftex,breaklinks=true,colorlinks=true]{hyperref}

\usepackage[utf8]{inputenc}

\usepackage{amssymb}%
\usepackage{epsfig}
\usepackage{tikz}
\newcommand{\journ}[5]
{\ifthenelse{\equal{#1}{pr}}{
Phys. Rev. {\bf #2}, \href{http://link.aps.org/abstract/PR/v#2/e#3}{#3} (#4)}
{\ifthenelse{\equal{#1}{prl}}{
Phys. Rev. Lett. {\bf #2}, \href{http://link.aps.org/abstract/PRL/v#2/e#3}{#3} (#4)}
{\ifthenelse{\equal{#1}{prb}}{
Phys. Rev. B {\bf #2}, \href{http://link.aps.org/abstract/PRB/v#2/e#3}{#3} (#4)}
{\ifthenelse{\equal{#1}{prd}}{
Phys. Rev. B {\bf #2}, \href{http://link.aps.org/abstract/PRB/v#2/e#3}{#3} (#4)}
{\ifthenelse{\equal{#1}{pra}}{
Phys. Rev. A {\bf #2}, \href{http://link.aps.org/abstract/PRA/v#2/e#3}{#3} (#4)}
{\ifthenelse{\equal{#1}{arxiv}}{
preprint \href{http://arxiv.org/abs/#2.#3}{arXiv:#2.#3}}
{\ifthenelse{\equal{#1}{rmp}}{
\rmp {\bf #2}, \href{http://link.aps.org/abstract/RMP/v#2/e#3}{#3} (#4)}
{\ifthenelse{\equal{#1}{cond-mat}}{
preprint \href{http://arxiv.org/abs/cond-mat/#2}{cond-mat/#2}}
{\ifthenelse{\equal{#1}{pre}}{
Phys. Rev. E {\bf #2}, \href{http://link.aps.org/abstract/PRE/v#2/e#3}{#3} (#4)}
{#1 {\bf #2}, #3 (#4)}}}}}}}}
}}

\newcommand{\journdoi}[6]{#1\ {\bf #2}, \href{http://dx.doi.org/#5}{#3} (#4)}

\newcommand{\la}{\langle}
\newcommand{\ra}{\rangle}
\begin{document}

\title{Phase transition in the R\'enyi-Shannon entropy of Luttinger liquids}

\author{Jean-Marie St\'ephan, Gr\'egoire Misguich and Vincent Pasquier}

\affiliation{Institut de Physique Th\'eorique,
CEA, IPhT, CNRS, URA 2306, F-91191 Gif-sur-Yvette, France.}

%%%%%%%%%%%%%%%%%%%%%%%%%%%%%%%%%%%%%%%%%%%%%%%%%
\begin{abstract}
%%%%%%%%%%%%%%%%%%%%%%%%%%%%%%%%%%%%%%%%%%%%%%%%%

The Rényi-Shannon entropy associated to  critical quantum spins chain 
with central charge $c=1$ is shown to have a phase transition at some 
value $n_c$ of the Rényi parameter $n$ which depends on the Luttinger 
parameter (or compactification radius R). 
Using a new replica-free formulation, the entropy is expressed as a 
combination of single-sheet  partition functions evaluated at $n-$ 
dependent values of the stiffness. The transition occurs when a vertex 
operator becomes relevant at the boundary. 
Our numerical results (exact diagonalizations for the XXZ and $J_1-J_2$ 
models) are in agreement with the analytical predictions: 
above $n_c=4/R^2$ the subleading and universal contribution to the 
entropy is $\ln(L)(R^2-1)/(4n-4)$ for open chains, and $\ln(R)/(1-n)$ 
for periodic ones (R=1 at the free fermion point). The replica approach 
used in previous works fails to predict this transition and turns out to 
be correct only for $n<n_c$.
From the point of view of two-dimensional 
Rokhsar-Kivelson 
states, the transition reveals a rich structure in the entanglement 
spectra.

%%%%%%%%%%%%%%%%%%%%%%%%%%%%%%%%%%%%%%%%%%%%%%%%%
\end{abstract}
%%%%%%%%%%%%%%%%%%%%%%%%%%%%%%%%%%%%%%%%%%%%%%%%%

\date{\today}


\maketitle
%%%%%%%%%%%%%%%%%%%%%%%%%%%%%%%%%%%%%%%%%%%%%%%
%%%%%%%%%%%%%%%%%%%%%%%%%%%%%%%%%%%%%%%%%%%%%%%
{\it Introduction}.---
%%%%%%%%%%%%%%%%%%%%%%%%%%%%%%%%%%%%%%%%%%%%%%%
%%%%%%%%%%%%%%%%%%%%%%%%%%%%%%%%%%%%%%%%%%%%%%%
The entanglement entropy (EE) 
has become an important tool to probe and characterize
many-body quantum states.
In one-dimensional (1d) systems, the celebrated logarithmic divergence \cite{hlw94,vlrk03,korepin04,cc04}
of the von Neumann entropy of a long segment provides an efficient way to measure the central charge of a critical spin chain.
In higher dimensions, EE can also be used to detect the presence of topological order \cite{kplw06}, which is
otherwise invisible to conventional local order parameters and correlation functions. Understanding
how scales the EE of large subsystems has also opened a route to algorithms able to simulate efficiently
 these strongly interacting systems in $d>1$ \cite{pepsmera}.

In this letter, we use a seemingly completely different entropy to probe the ground-state of quantum spin chains,
 the {\it Shannon entropy} (or configuration entropy).
For a normalized state $|\psi\ra$ and a Rényi index $n>0$ it is defined as :
\begin{eqnarray}\label{eq:renyishannon}
 S_n&=&\frac{1}{1-n}\ln \left(\sum_i p_i^n\right) \;\;\;p_i=| \la i|\psi\ra |^2
\end{eqnarray}
where the states $|i\ra$ form a basis of the Hilbert space. The basis
states are chosen to be products of local states, and the Ising configurations 
($|i\ra=|\uparrow\uparrow\downarrow\cdots\ra$, etc.)
are the natural choice for the spin-$\frac{1}{2}$ systems we consider.
This entropy was originally introduced as a mean to evaluate the EE in some particular two-dimensional (2d) states -- the so called Rokhsar-Kivelson (RK) wave functions -- and has
allowed to investigate in details the EE at some  some 2d conformal quantum critical points \cite{stephan09,zbm11}.
We will interpret our results in terms of 2d entanglement spectra at the end, but we
will otherwise  mostly concentrate on
the 1d spin chain point of view.

A particularly interesting situation is that of the spin-$\frac{1}{2}$ XXZ chain:
\begin{equation}\label{eq:xxz}
 \mathcal{H}=\sum_i \left(S^x_i\cdot S^x_{i+1} + S^y_i\cdot S^y_{i+1}\right) + \Delta \sum_i S^z_i\cdot S^z_{i+1}
\end{equation}
For a chain of length $L$, $S_n$ has a leading term proportional to $L$ followed by universal subleading contributions.
If the chain is {\it periodic}, the first subleading term is of order $O(1)$ and was shown to be \cite{stephan09,oshikawa10} :
\begin{equation}
 S_n^{\rm periodic}= (\cdots) L +  \ln\left(R\right)-\frac{\ln n}{2(n-1)}, \label{eq:Sn}
\end{equation}
where $R$ is the compactification radius, related to the anisotropy $\Delta$ of the XXZ Hamiltonian ($R(\Delta)^2=2-\frac{2}{\pi}\arccos(\Delta)$).
 With {\it open} boundary conditions the first subleading correction is also universal and takes the form of a logarithm of the length $L$ of the chain \cite{zbm11}:
\begin{equation}\label{eq:cp}
 S_n^{\rm open}= (\cdots) L -\frac{1}{4}\ln(L),
\end{equation}
which, in the 2d-RK language, confirms at $c=1$ the arguments developed in \cite{fm06}
(from now on the term proportional to $L$ will be omitted).

In this letter we however show that these results are only correct below a critical value $n_c$ of the Rényi parameter.
Using a replica-free formulation of the problem we prove that the Rényi parameter $n$
effectively modifies the compactification radius of the chain (in a  sense to be defined latter),
and that a phase transition takes place at $n=n_c$ when a (boundary) vertex operator becomes relevant.
A central result concerns the location of this transition and the behavior of the entropy above $n_c$:
\begin{eqnarray}
 n_c &=& d^2 /R^2 \label{eq:nc}\\
 S_{n>n_c}^{\rm periodic} &=&\frac{1}{n-1}\left(n\ln R - \ln d \right)  \label{eq:Pnc} \\
 S_{n>n_c}^{\rm open} &=&\ln(L)\frac{n}{n-1}\left(\frac{R^2}{4}-\frac{1}{4}\right) , \label{eq:Onc}		    
\end{eqnarray}
where $d$ is the degeneracy of the Ising configuration with the highest probability $p_{\rm max}$ in the ground state.
At zero magnetization this configuration is $d=2$-fold degenerate:
\begin{equation}
|i_{\rm max}\rangle=|\!\!\uparrow\downarrow\uparrow \downarrow\cdots \uparrow\downarrow \ra \;\;\textrm{  or  }\;\;
|i_{\rm max}\rangle=|\!\!\downarrow\uparrow\downarrow \uparrow\cdots \downarrow\uparrow \ra.
\end{equation}

%%%%%%%%%%%%%%%%%%%%%%%%%%%%%%%%%%%%%%%%%%%%%%%
%%%%%%%%%%%%%%%%%%%%%%%%%%%%%%%%%%%%%%%%%%%%%%%
{\it Compact free field}.---
%%%%%%%%%%%%%%%%%%%%%%%%%%%%%%%%%%%%%%%%%%%%%%%
%%%%%%%%%%%%%%%%%%%%%%%%%%%%%%%%%%%%%%%%%%%%%%%
To understand this transition we adopt the 1+1 dimensional (Euclidian) point of view.
 At long distances, the model is described by a compactified ``height field'' $h(x,\tau)$ with Gaussian probabilities:
\begin{eqnarray}
S[h]=\frac{\kappa}{4\pi}\int dx d\tau ({\bf \nabla} h)^2 \label{eq:S} \\
 \mathcal Z= \int \mathcal{D}[h] \; \exp\left(-S[h]\right). \label{eq:Z} 
\end{eqnarray}
where $\kappa$ is the stiffness, $r$ the compactification radius ($h \equiv h+2\pi r$)
and the physical Luttinger parameter (which fixes the decay exponents of the correlations functions) is $R=\sqrt{2\kappa} r$.
For periodic (resp. open) chains $h$ lives on an infinitely long cylinder (resp. strip) of perimeter (resp. width) $L$.
In this language, the microscopic configurations $|i\ra$
are replaced by configurations $\phi(x)=h(x,\tau=0)$ of the height field at $\tau=0$.
To evaluate the probability $p[\phi]$ we decompose the field $h$ into an harmonic function $h^\phi$ which satisfies the boundary condition $h^\phi_{\tau=0}=\phi$,
 and a ``fluctuating'' part $\delta h$ satisfying a Dirichlet boundary condition: $h=h^\phi + \delta h$ 
\footnote{The winding numbers $W$ need not be summed over since only $W=0$ contributes in the limit of infinite length in the imaginary time direction.
}.
Exploiting the Gaussian form of the action and  $\Delta h_\phi=0$, the ``classical'' and ``fluctuating'' part decouple:
$
S[h]=S[h^\phi]+S[\delta h]
$ and we get
\begin{equation}
 p[\phi]=\exp(- S[h^\phi]) \frac{\mathcal{Z}^D}{\mathcal{Z}}, \label{eq:pphi}
\end{equation}
where  $\mathcal{Z}^D$ is the partition function of the whole cylinder (resp. strip) with a Dirichlet defect
line at $\tau=0$ \footnote{$\mathcal{Z}^D$ is thus the square of the partition function a half-infinite cylinder (strip).}.
Now $p[\phi]$ is raised to the (possibly non-integer) power $n$
\begin{equation}
 p[\phi]^n=\exp(-n S[h^\phi]) \left (\frac{\mathcal{Z}^D}{\mathcal{Z}} \right)^n, \label{eq:pn}
\end{equation}
and make the observation that $\exp(-n S[h^\phi])$ is the Boltzmann weight in a system where the stiffness
$\kappa$ has been replaced by $\kappa'=n\kappa$. From now on we explicitly keep track of the value of the stiffness $\kappa$ (as an index) and write:
\begin{equation}
 \exp(-n S_\kappa[h^\phi])= \exp(-S_{n\kappa}[h^\phi])=p_{n\kappa}[\phi] \frac{\mathcal Z_{n\kappa}}{Z^D_{n\kappa}}.
\end{equation}

So, Eq.~\ref{eq:pn} can be written as
\begin{equation}
 \left(p_{\kappa}[\phi]\right)^n=p_{n\kappa}[\phi] \left(\frac{\mathcal Z_{n\kappa}}{Z^D_{n\kappa}}\right)\left ( \frac{\mathcal{Z}^D_\kappa}{\mathcal{Z}_\kappa} \right)^n .
\end{equation}
Inserting this result in Eq.~\ref{eq:renyishannon} and using the fact that the probabilities $p_{n\kappa}[\phi]$ are
 normalized, we get the main result of this section:
\begin{eqnarray}
 S_n&=&\frac{1}{1-n}\left[
\ln\left(\frac{\mathcal Z_{n\kappa}}{\mathcal Z^D_{n\kappa}}\right) -n \ln\left(\frac{\mathcal Z_\kappa}{Z^D_\kappa}\right)
\right]
\label{eq:RnZZD}
\end{eqnarray}
If we assume $2n=p$ to be an integer the partition function $Z(n)= \sum p_i^n$ has a natural interpretation in terms of
 $p$ half-infinite systems glued together at their edge, forming a ``book'' with $p$ sheets \cite{smp10}.
The derivation above, however, never assumes $2n$ to be an integer and is therefore different from the previous derivations
involving a replica trick \cite{oshikawa10,zbm11}.
$S_n$ has been reduced to ratios of standard partition functions with respectively free and Dirichlet boundary conditions
at the boundary between upper ($\tau>0$) and lower ($\tau<0$) parts of the cylindric (or strip). The complications associated
 to $n$-sheeted surfaces \cite{oshikawa10,zbm11} have been avoided.
For a compactified free field and the cylinder geometry, the ratio
 $g_D^2=\mathcal Z_{\kappa}^D/\mathcal Z_{\kappa}$ is a well known ``$g$-factor'' \cite{al91,fsw94}:
\begin{equation}
 g_D^2=R^{-1}=(2\kappa r^2)^{-1/2}.\label{eq:gd2}
\end{equation}
Combining Eq.~\ref{eq:RnZZD} and \ref{eq:gd2} we find:
\begin{eqnarray}
S_n&=&\ln R-\frac{\ln n}{2(n-1)}, \label{eq:SnP}
\end{eqnarray}
in agreement with Refs.~\cite{stephan09,oshikawa10,hsu10b,zbm11}.
For the strip geometry, we also recover the $n$-independent result of Eq.~\ref{eq:cp} by applying
the Cardy-Peschel formula \cite{cp88} to Eq.~\ref{eq:RnZZD}, with four angles $\gamma=\pi/2$.

%%%%%%%%%%%%%%%%%%%%%%%%%%%%%%%%%%%%%%%%%%%%%%%
%%%%%%%%%%%%%%%%%%%%%%%%%%%%%%%%%%%%%%%%%%%%%%%
{\it Boundary phase transition}.---
%%%%%%%%%%%%%%%%%%%%%%%%%%%%%%%%%%%%%%%%%%%%%%%
%%%%%%%%%%%%%%%%%%%%%%%%%%%%%%%%%%%%%%%%%%%%%%%
Eq.~\ref{eq:RnZZD} is a combination of two $g$-factors and therefore probes the boundary of the system (the ``bookbinding'' in the book picture).
The action at the lattice scale is not strictly Gaussian and other terms respecting the lattice symmetry and the periodicity $h\equiv h+2\pi r$
are present, including vertex operators of the type $\cos(\frac{d}{r}h)$.
At the boundary such an operator renormalizes to zero in the long distance limit if $d^2 > 2\kappa r^2 $ \cite{coleman75}.
Otherwise it would lock the field to a flat configuration with degeneracy $d$.
But the Eq.~\ref{eq:RnZZD} cannot be valid anymore in this locked/massive phase since its derivation assumes Gaussian probabilities.
A boundary phase transition therefore takes place when the most relevant vertex operator become marginal in presence of a stiffness $\kappa'=n\kappa$ 
\footnote{A locking transition in the {\it bulk}
is expected at $d^2 = 4\kappa r^2$. It however turns out that the latter does not affect Eq.~\ref{eq:RnZZD}, as confirmed by our numerical results.
In the ``book`` picture  and sufficiently far from the boundary it is indeed clear that all the pages (with a {\it fixed} stiffness $\kappa$)
 remain gapless in the bulk,  whatever $n$.}.
This gives the critical value of the Rényi index:
\begin{equation}
  n_c = \frac{d^2} {2 \kappa r^2}.
\end{equation}
In the case of the antiferromagnetic XXZ chain the
two Ising  configurations $|\uparrow\downarrow\uparrow \downarrow\cdots\ra$ and $|\downarrow\uparrow\downarrow\uparrow\cdots\ra$ become the ground-states when $\Delta\to\infty$,
which shows that an operator with two minima is present at the microscopic level.
Taking $2\kappa r^2 = R(\Delta)$ and $d=2$, we find Eq.~\ref{eq:nc}. We also note that the same argument applies to the $J_1-J_2$ (or ``zig-zag'') chain,
and to any Luttinger liquid phase with umklapp terms $\sim \cos(2h/r)$.\footnote{Such a behavior has already been observed in a dimer problem \cite{stephan09}.}

In this locked phase, the universal contribution to $S_n$ is given by these  $d$ configurations only,
so that
\begin{equation}
 S_{n>n_c}=\frac{1}{1-n}\ln\left[d (p_{\rm max})^n\right].
\end{equation}
In the periodic case, $p_{\rm max}$ simply corresponds to $g_D^2=\mathcal Z^D/\mathcal Z$ and we recover  Eq.~\ref{eq:Pnc}.

\begin{figure}
 \includegraphics[width=7cm]{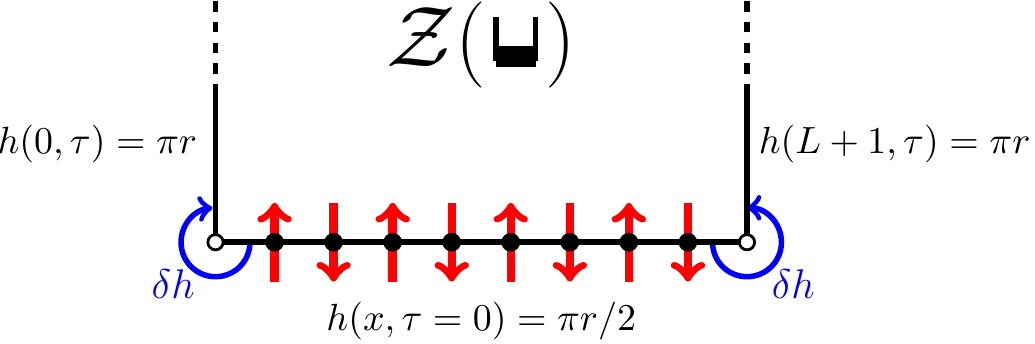}
\caption{(Color online) Height shift $\delta h=\pi r/2$ for the semi-infinite strip with bottom boundary condition $|i_{\rm max}\rangle$.}
\label{fig:shift}
\end{figure}

The open chains turn out to be even more interesting. The universal contribution to $S_{n>n_c}$ is encoded in
\begin{equation}\label{eq:pmax}
 p_{\rm max}
=\lim_{\tau \to \infty}
\frac{\left|\langle s|e^{-\tau H}|i_{\rm max}\rangle\right|^2}
{\langle s|e^{-2 \tau H}|s\rangle}
=
\frac{\mathcal{Z}(
\begin{tikzpicture}
\draw[thick] (0,-0.025) -- (0,0.2);
\draw[thick] (0.2,-0.025) -- (0.2,0.2);
\draw[line width=3pt] (0,0) -- (0.2,0);
\end{tikzpicture}
)^2
}
{\mathcal{Z}(
\begin{tikzpicture}
\draw[thick] (0,-0.15) -- (0,0.15);
\draw[thick] (0.2,-0.15) -- (0.2,0.15);
\end{tikzpicture}
)
}
,
\end{equation}
where $\mathcal{Z}(
\begin{tikzpicture}
\draw[thick] (0,-0.025) -- (0,0.2);
\draw[thick] (0.2,-0.025) -- (0.2,0.2);
\draw[line width=3pt] (0,0) -- (0.2,0);
\end{tikzpicture}
)$ 
 is the partition function of a semi-infinite strip with bottom boundary condition $|i_{\rm max}\rangle$ (see Fig.~\ref{fig:shift}).
 The ratio in Eq.~\ref{eq:pmax} is similar, but not identical
 to the one used below the transition ($n<n_c$). As before, four corners with angle $\pi/2$ will contribute
 to $-\ln p_{\rm max}$ by a logarithm: $-\frac{1}{4} \ln L$. However, the configuration with highest probability
 does not exactly correspond to Dirichlet in the continuum limit since there is a {\it height shift} $\delta$
 between the vertical edges of the strip, and the horizontal boundary (see Fig.~\ref{fig:shift}). It can be treated
 by subtracting an harmonic function $h_{\delta}(x,\tau)=(2\delta/\pi) \arg (x+i\tau)$, equal to $0$ on the horizontal boundary
 $\tau=0$ and $\delta$ on the vertical boundary at $x=0$. The resulting contribution to the free energy is
\begin{equation}\label{eq:energy_shift}
 \delta F \sim \frac{\kappa \delta^2}{2\pi^2} \ln L.
\end{equation}
The value shift $\delta$ can be obtained using, for instance, a bosonization approach.
The free boundary condition for the spins at the end of the open chain corresponds to Dirichlet for the free field
and we have to set $h(x=0,\tau)=h(x=L+1,\tau)=\pi r$ to insure vanishing spin operators at both ends \cite{egg92}. 
Then, the continuum limit of the configuration $|i_{\rm max}\ra$
 corresponds to locking $h(x,\tau=0)$ to the minima of the umklapp term $\cos(2 h/r)$ \cite{Afflecklh}, which has
two degenerate minima in $h=\pi r/2$ and $3\pi r/2$. In both cases the height difference between the $x=0$
boundary and that at $\tau=0$ is $\delta=\pi r/2$.
Summing up the contributions coming
 from Eq.~\ref{eq:energy_shift} and from the Cardy-Peschel term, we recover our main result Eq.~\ref{eq:Onc}.

\begin{figure}
 \includegraphics[width=8.7cm]{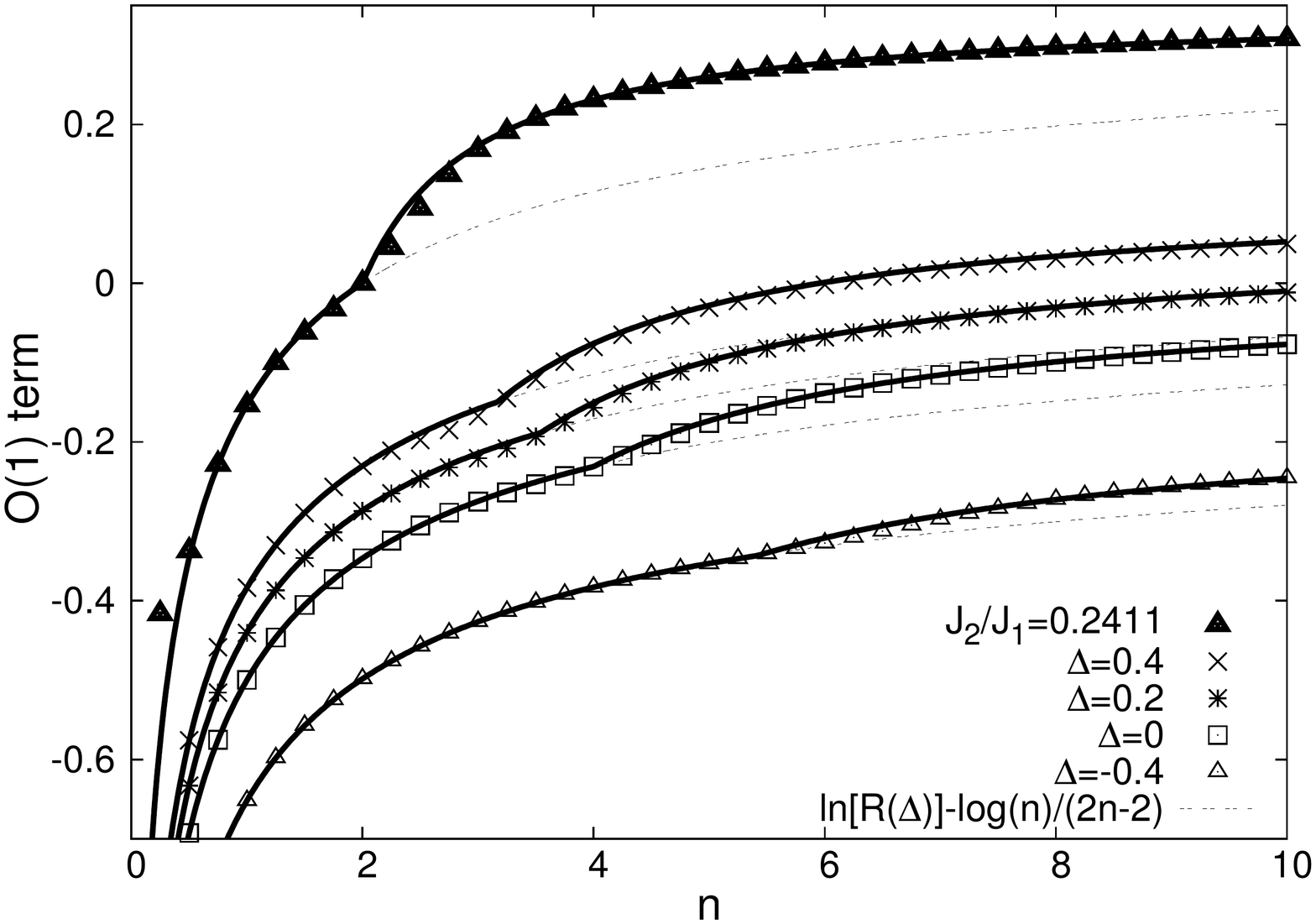}
 \caption{Constant term in the Shannon-Rényi entropy of periodic  XXZ and $J_1-J_2$ chains for different
 values of  $\Delta$ and at the critical point $J_2/J_1=0.2411$. Each  point comes from fitting
 the data for $L=20,22,24,26$ and $28$ to  $aL+b+c/L+d/L^2$. Fat lines: theoretical prediction (Eq.~\ref{eq:RnZZD}).
Eq.~\ref{eq:Sn} is also plotted above $n_c$ (dashed lines) for comparison.}
  \label{fig:XXZ_P}
\end{figure}

%%%%%%%%%%%%%%%%%%%%%%%%%%%%%%%%%%%%%%%%%%%%%%%
%%%%%%%%%%%%%%%%%%%%%%%%%%%%%%%%%%%%%%%%%%%%%%%
{\it Numerical simulations}.---
%%%%%%%%%%%%%%%%%%%%%%%%%%%%%%%%%%%%%%%%%%%%%%%
%%%%%%%%%%%%%%%%%%%%%%%%%%%%%%%%%%%%%%%%%%%%%%%
So far, for periodic chains,  only the case  $n=1$  has been investigated numerically \cite{stephan09}. Fig.~\ref{fig:XXZ_P}
shows the full $n$ dependence of $S_n$ (constant term extracted by fitting the finite-size data)
for different values of $\Delta$. Although the system sizes are very small ($L\leq 28$)
there is a good agreement with the theoretical predictions,  including the change of behavior at the predicted
value of $n_c$ (which depends on $\Delta$). It is only close to the Heisenberg point ($\Delta=1$) and
above $n_c$ that the finite-size results deviate from Eq.~\ref{eq:Onc}. We attribute these enhanced finite-size effects to
the marginal operators present at the $SU(2)$ symmetric point. To circumvent this difficulty
 we also studied the $J_1-J_2$ chain at the critical value $J_2/J_1 \sim 0.2411$ \cite{on92}, which has the same radius $R=\sqrt{2}$
 but where finite-size effects are much smaller. The data again agrees well with our prediction.

\begin{figure}
 \includegraphics[width=8.7cm]{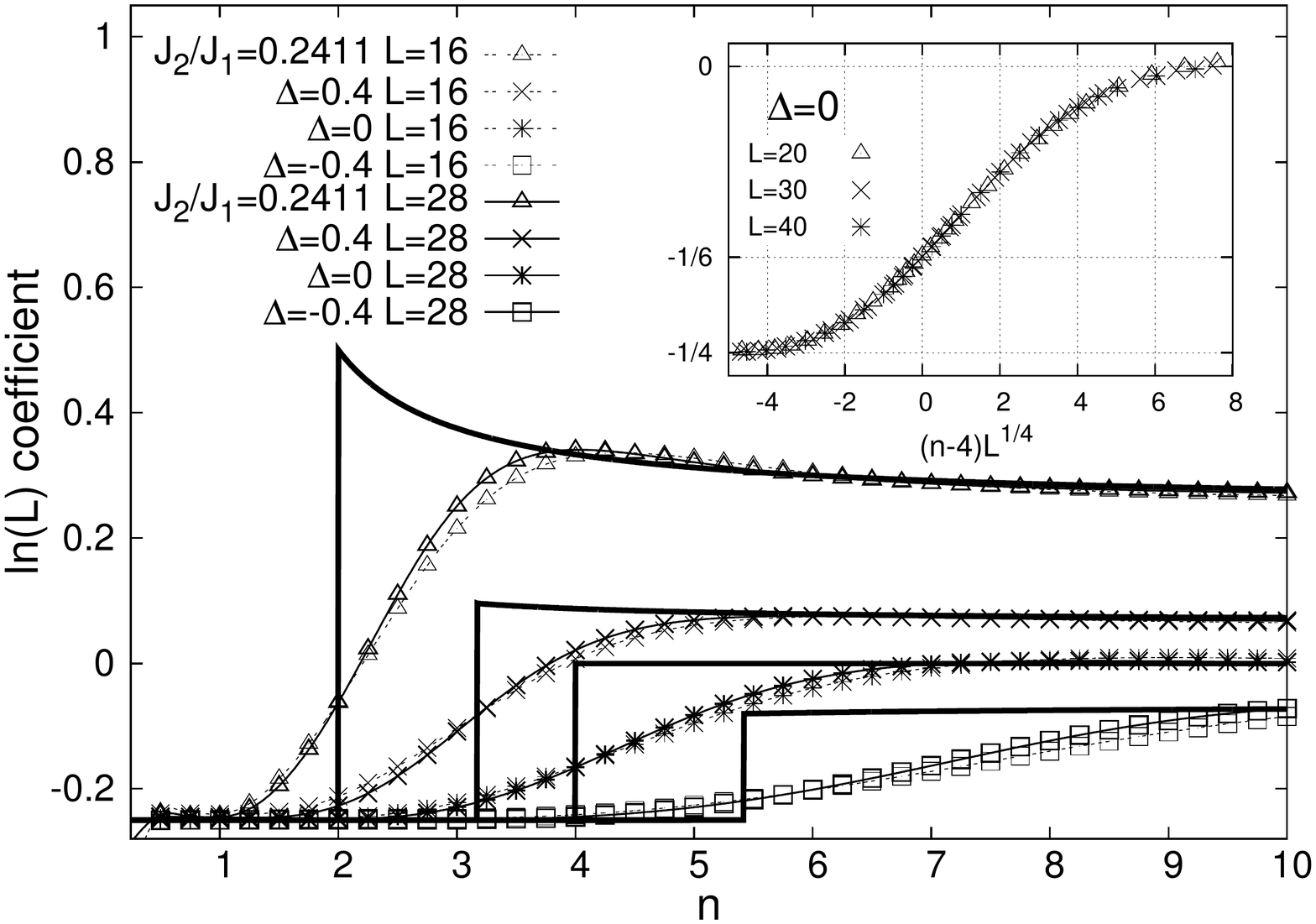}
 \caption{Logarithmic term in the Shannon-Rényi entropy of open  XXZ and $J_1-J_2$  chains, extracted using a fit $aL+b\ln L+c+d/L$ to four consecutive even systems sizes ($L,\cdots,L-6$) with  $L=16$ and $L=28$.
Fat lines: Eq.~\ref{eq:RnZZD}. Inset: scaling close to $n_c(\Delta=0)=4$.}
\label{fig:XXZ_O}
\end{figure}

Open chains were investigated in Ref.~\cite{zbm11} for various values of
$n$ and $\Delta$, but the transition at $n_c$ was overlooked. In Fig.~\ref{fig:XXZ_O}
we see a clear tendency for the logarithmic term to approach the theoretical curves, although the finite size data
are still far from the thermodynamic limit. Interestingly, the entropies curves extracted from
different system sizes cross in the immediate vicinity of the predicted value of $n_c$. This lead us to conjecture that
the coefficient of the logarithm may take a universal value at the transition point which only depends on the compactification radius $R$.
At $\Delta=0$  we pushed the numerics  to $L=40$ spins (see below) and
 the inset of Fig~\ref{fig:XXZ_O} shows a collapse of the data for different system sizes onto a single curve in the vicinity
 of $n_c=4$. This indicates a slow ($\sim L^{-1/4}$) but steady convergence towards a step function. We also get
 $S_{n=n_c}^{\rm open}=-(1/6) \ln L$ with a high  precision at $\Delta=0$, although we have no theoretical understanding of this value.


%%%%%%%%%%%%%%%%%%%%%%%%%%%%%%%%%%%%%%%%%%%%%%%
{\it $\Delta=0$ in open chains}.---
%%%%%%%%%%%%%%%%%%%%%%%%%%%%%%%%%%%%%%%%%%%%%%%
This corresponds to free fermions
and each probability $p_i$ is a determinant (Wick's theorem).
Denoting by $\{x_k\}$ the positions of the up spins (fermions) 
 and  setting $\theta_k=\pi x_k/(L+1)$ we have:
 \begin{eqnarray}
p_i&=& p\left(\{\theta_k\}\right)=\left(\frac{2}{L+1}\right)^{L/4}\det_{1\leq j,k\leq n}\left(\sin \left[j\theta_k\right]\right)
\end{eqnarray}
 This determinant can be computed exactly \cite{kratt}:
\begin{eqnarray}\nonumber
p_i&=&\prod_{j=1}^{L/2} \frac{2 \sin^2 \theta_j}{L+1}\prod_{k>j}16 \sin^2 \left(\frac{\theta_j-\theta_k}{2}\right) \sin^2 \left(\frac{\theta_j+\theta_k}{2}\right)
\end{eqnarray}
and allows to go get $p_{\rm max}$:
$
 p_{\max}=2^{-L/2}.
$
The absence of a logarithmic term in $S_{n \to\infty}\sim-\ln p_{\rm max}$ is  consistent with Eq.~\ref{eq:Onc}, because
the contribution from the height shift
exactly compensates for the Cardy-Peschel term at $R(\Delta=0)=1$. This 
also rigorously confirms at $\Delta=0$ the value of $\delta$.
We also checked the validity of Eq.~\ref{eq:Onc}
for the square lattice quantum dimer (2d-RK) wave function, 
 for which $d=1$. In this case
 the microscopic height representation of dimer coverings allows to obtain $\delta$ exactly at the lattice level.

%%%%%%%%%%%%%%%%%%%%%%%%%%%%%%%%%%%%%%%%%%%%%%%
{\it Conclusion}.---
%%%%%%%%%%%%%%%%%%%%%%%%%%%%%%%%%%%%%%%%%%%%%%%
The Rényi-Shannon entropy of critical 1d systems does not only give a 
simple access to the Luttinger parameter of the system, 
it also shows that taking {\it powers} of the wave function 
gives rise to a whole line of critical points ending at a phase 
transition to an ordered state. In fact, going back 
to the  2d RK point of view, this transition reveals a rich structure in 
the entanglement spectrum $\{E_i\}$ of these 2d wave-functions. 
In some appropriate geometry \cite{stephan09} the probabilities $p_i$ 
are nothing else but the eigenvalues of the RK reduced density matrix 
and the probabilities directly give the entanglement spectrum $E_i=-\ln 
p_i$. The phase transition at $n_c$ shows that the spectrum has two 
sharply distinct regions in the thermodynamic limit: at high ``energy'' 
(small inverse ``temperature'' $n$) the universal contributions to the 
probabilities are Gaussian whereas at low energy (large $n$) they are 
dominated by the ordered configurations. 
This contrasts with the fractional quantum Hall situation where the 
universal part of the entanglement spectrum only exist at low $E_i$ \cite{hlc}.
These findings may open new directions in the study of quantum critical 
wave-functions -- in 1d and in higher dimensions -- 
as well as their connection to boundary critical phenomena and 
$g$-factors in particular.

%%%%%%%%%%%%%%%%%%%%%%%%%%%%%%%%%%%%%%%%%%%%%%%
{\it Aknowledgments}.---
%%%%%%%%%%%%%%%%%%%%%%%%%%%%%%%%%%%%%%%%%%%%%%%
We are  indebted to Edouard Boulat for suggesting us the bosonization argument to compute $\delta$,
 and thank Hubert Saleur for insightful discussions.

%%%%%%%%%%%%%%%%%%%%%%%%%%%%%%%%%%%%%%%%%%%%%%%
%%%%%%%%%%%%%%%%%%%%%%%%%%%%%%%%%%%%%%%%%%%%%%%%%

%%%%%%%%%%%%%%%%%%%%%%%%%%%%%%%%%%%%%%%%%%%%%%%%%


\begin{thebibliography}{99}
%%%%%%%%%%%%%%%%%%%%%%%%%%%%%%%%%%%%%%%%%%%%%%%%%
%%%%%%%%%%%%%%%%%%%%%%%%%%%%%%%%%%%%%%%%%%%%%%%

\bibitem{hlw94}
 C. Holzhey, F. Larsen, and F. Wilczek,
 \journdoi{Nucl. Phys. B}{424}{443}{1994}{Geometric and renormalized entropy in conformal field theory}{10.1016/0550-3213(94)90402-2}.
\bibitem{vlrk03}
 G. Vidal, J. I. Latorre, E. Rico, and A. Kitaev,
 \journ{prl}{90}{227902}{2003}{Entanglement in Quantum Critical Phenomena}.
\bibitem{korepin04}
 V. E. Korepin,
 \journ{prl}{92}{096402}{2004}{Universality of Entropy Scaling in One Dimensional Gapless Models}.
\bibitem{cc04}
P. Calabrese and J. Cardy,
\journdoi{J. Stat. Mech.}{}{P06002}{2004}{10.1088/1742-5468/2004/06/P06002}{Entanglement entropy and quantum field theory}.
\bibitem{kplw06}
A. Kitaev and J. Preskill, \journ{prl}{96}{110404}{2006}{Topological Entanglement entropy}; 
M. Levin and X.-G. Wen, \journdoi{ibid}{}{110405}{2006}{10.1103/PhysRevLett.96.110404}{Detecting Topological order in a Ground State Wave Function}.
\bibitem{pepsmera}
J. I. Cirac and F. Verstraete, \journdoi{J. Phys. A: Math. Theor.}{42}{504004}{2009}{10.1088/1751-8113/42/50/504004}{Renormalization and tensor products states
 in spin chains and lattices}.
\bibitem{stephan09}
J.-M. St\'ephan {\it et al.}, 
\journ{prb}{80}{184421}{2009}{Shannon and entanglement entropies of one- and two-dimensional critical wave functions}.

\bibitem{zbm11}
M. P. Zaletel, J. H. Bardarson, J. E. Moore,
\journ{arxiv}{1103}{5452}{2011}{Logarithmic terms in entanglement entropies of 2D quantum critical points and Shannon entropies of spin chains}.

\bibitem{oshikawa10}
M. Oshikawa,
\journ{arxiv}{1007}{3739}{2010}{Boundary Conformal Field Theory and Entanglement Entropy in Two-Dimensional Quantum Lifshitz Critical Point}.

\bibitem{fm06}
E. Fradkin and J.E. Moore,
 \journ{prl}{97}{050404}{2006}{Entanglement Entropy of 2D Conformal Quantum Critical Points: Hearing the Shape of a Quantum Drum}.
 
\bibitem{smp10}
J.-M. St\'ephan, G. Misguich, and V. Pasquier,
\journ{prb}{82}{125455}{2010}{R\'enyi entropy of a line in two-dimensional Ising models}.



\bibitem{al91}
I. Affleck and A. W. W Ludwig, \journ{prl}{67}{161}{1991}{}.
\bibitem{fsw94}
P. Fendley, H. Saleur and N. Warner, 
\journdoi{Nucl. Phys. B}{430}{577}{1994}{Exact solution of a massless scalar field with a relevant boundary interaction}
{10.1016/0550-3213(94)90160-0}.

\bibitem{hsu10b}
B. Hsu and E. Fradkin,
\journdoi{J. Stat. Mech}{}{P09004}{2010}{10.1088/1742-5468/2010/09/P09004}{Universal behavior of entanglement in 2D quantum critical dimer models}.

\bibitem{cp88}
J. Cardy and I. Peschel, \journdoi{Nucl. Phys. B}{300}{377}{1988}{10.1016/0550-3213(88)90604-9}{Finite-size dependence of the free energy in two-dimensional critical systems}.

\bibitem{coleman75}
S. Coleman,
\journ{prd}{11}{2088}{1975}{Quantum sine-Gordon equation as the massive Thirring model}.


\bibitem{egg92}
S. Eggert and I. Affleck, \journ{prb}{46}{10866}{1992}{Magnetic impurities in half-integer-spin Heisenberg antiferromagnetic chains}
\bibitem{Afflecklh}
I. Affleck, {\it Fields, Strings and Critical Phenomena}, p563-640, proceedings of Les Houches Summer School, 1988.
\bibitem{on92}
K. Okamoto and K. Nomura,
\journdoi{Phys. Lett. A}{169}{433}{1992}{Fluid-dimer critical point in S = Image  antiferromagnetic Heisenberg chain with next nearest neighbor interactions}{10.1016/0375-9601(92)90823-5}.

\bibitem{kratt}
C. Krattenthaler, preprint \href{http://arxiv.org/abs/math/9902004}{math/9902004}.
\bibitem{hlc}
H. Li and F. D. M Haldane, \journ{prl}{101}{010504}{2008}{Entanglement spectrum as a Generalization of Entanglement Entropy:
 Identification of Topological order in Non-Abelian Fractional Quantum Hall Effect States};
R. Thomale, A. Sterdyniak, N. Regnault and B. A. Bernevig,
\journ{prl}{104}{180502}{2010}{The entanglement gap and a new principle of adiabatic continuity}.



%%%%%%%%%%%%%%%%%%%%%%%%%%%%%%%%%%%%%%%%%%%%%%%%%
\end{thebibliography}
\end{document}